\documentclass[aps,prl,reprint,superscriptaddress,nobibnotes]{revtex4-2}

\usepackage{graphicx}% Include figure files
\usepackage{color}
\usepackage[version=3]{mhchem}
\usepackage{CJKutf8}

\begin{document}
\begin{CJK*}{UTF8}{gbsn}
%\preprint{}{\tiny }

\title{Monitoring Vibrational Evolution in Jahn-Teller Effect by Raman Images}

%\author{}
%\thanks{}
%\affiliation{}


\author{Hai-Zhen Yu}
\affiliation{Shandong Province Key Laboratory of Medical Physics and
	Image Processing Technology, School of Physics and Electronics, Shandong
	Normal University, Jinan 250014, P.R. China.}

\author{Dingwei Chu}
\affiliation{Shandong Province Key Laboratory of Medical Physics and
	Image Processing Technology, School of Physics and Electronics, Shandong
	Normal University, Jinan 250014, P.R. China.}

\author{Yuanzhi Li}
\affiliation{Shandong Province Key Laboratory of Medical Physics and
	Image Processing Technology, School of Physics and Electronics, Shandong
	Normal University, Jinan 250014, P.R. China.}

\author{Li Wang}
\affiliation{Shandong Province Key Laboratory of Medical Physics and
	Image Processing Technology, School of Physics and Electronics, Shandong
	Normal University, Jinan 250014, P.R. China.}

\author{Yuzhi Song}
\affiliation{Shandong Province Key Laboratory of Medical Physics and
    Image Processing Technology, School of Physics and Electronics, Shandong
    Normal University, Jinan 250014, P.R. China.}

\author{Sai Duan}
\email{duansai@fudan.edu.cn}
\affiliation{Collaborative Innovation Center of Chemistry for Energy
	Materials, Shanghai Key Laboratory of Molecular Catalysis and Innovative
	Materials, MOE Key Laboratory of Computational Physical Sciences,
	Department of Chemistry, Fudan University, Shanghai 200433, P.R. China}
\affiliation{Hefei National Laboratory, Hefei 230088, P.R. China}

\author{Chuan-Kui Wang}
\email{ckwang@sdnu.edu.cn}
\affiliation{Shandong Province Key Laboratory of Medical Physics and
Image Processing Technology, School of Physics and Electronics, Shandong
Normal University, Jinan 250014, P.R. China.}

\author{Zhen Xie}
\email{xiezhen@sdnu.edu.cn}
\affiliation{Shandong Province Key Laboratory of Medical Physics and
	Image Processing Technology, School of Physics and Electronics, Shandong
	Normal University, Jinan 250014, P.R. China.}

\date{\today}

\begin{abstract}
The Jahn-Teller effect (JTE) reduces the geometrical symmetry of a system with
degenerate electronic states via vibronic coupling, playing a pivotal
role in molecular and condensed systems. In this Letter, we propose that
vibrational resolved tip-enhanced Raman scattering images can visualize
the vibrational evolutions in JTE in real space. Taking an
experimentally viable single zinc phthalocyanine (ZnPc) molecule as a
proof-of-principle example, not only the degenerate vibrational
splitting but also the overlooked vibration mixing caused by the JTE in
its anionic form can be straightforwardly characterized by Raman images.
Leveraging Raman images, the controllable configuration of JTE distortion
with partial isotopic substitution could be further identified. These
findings establish a practical protocol to monitor the detailed
vibrational evolutions when a single molecule experiences JTE, opening a
door for visualization of spontaneous symmetry breaking in molecular and
solid-state systems.
\end{abstract}

%\keywords{}

\maketitle
\end{CJK*}

\section{}

The Jahn-Teller effect (JTE)\cite{jahn-taller1937,chem.rev.2021}
describes a fundamental phenomenon that a degenerate system spontaneously
reduces its geometry symmetry via the vibronic coupling to stabilize the
total energy\cite{1975pnas,2019PhysRevX.,1958PRSLA}. JTE represents a
general mechanism of spontaneous symmetry breaking, which is responsible for
various important observations across multiple disciplines, including
spectroscopy, molecular and solid-state physics, stereochemistry, and
materials science\cite{APL2021,11PRB,15PRB,21PRL, nature2023,17Nat.Commun.,Du2025PRL}.
In practice, JTE can be efficiently induced by injecting
or removing an electron from a degenerate electronic 
state\cite{charge,2023angew, 2022afm}, particularly in systems containing transition metal
ions\cite{2022JACS,19PRB,11PRB}. In this context, scanning probe
microscopy (SPM) techniques provide a unique means to precisely control
JTE in real space\cite{JT2019nature,JT2019PRL}.

Despite the success of SPM techniques in the visualization of JTE,
most investigations were focused on imaging the symmetry breaking of
electronic states\cite{science2005,PRB2010stmjte,1999PRLjt,2018PRL-jt,STMjte2024PRL}.
Exclusive investigation on the electronic degrees of freedom in JTE has
led to a severe lack of understanding of its vibrational aspect, which
results in potentially questionable assignments of accompanying
frequency splitting, the overlooked mixing of vibrational wavefunctions upon
symmetry reduction, and the unexplored possibility of controlling JTE
distortions through atomic displacements.
Indeed, only few experiments studied the vibrational feature of JTE by SPM. For
instance, by leveraging the tip-enhanced Raman scatting (TERS)\cite{CPL2000,APL2000,OP2000,JP2000},
recent measurements have captured the spectral variations associated with JTE
in a negatively charged zinc phthalocyanine (\ce{ZnPc-})\cite{2023angew}, where
frequency evolution was highlighted. To date, there have been no reports
on imaging vibrational symmetry breaking, another essential aspect of
JTE, although it is possible owing to the high spatial resolution of
TERS\cite{2013nature,2021science,duan2024csr,2023duanNat.Photon.,li2020nano.tech.}.

In this Letter, taking the experimentally feasible ZnPc molecule as a
proof-of-principle example, we theoretically propose that TERS images
provide a powerful means to monitor all the details of the
vibrational evolution in JTE distortion. We demonstrate that,
not only the splitting of degenerate vibrations but also the mixing of
non-degenerate vibrations belonging to different irreducible
representations of the high-symmetry point group, can be affirmatively
observed in TERS images. For the latter case, we reveal that the mixing
strength is determined by both the energy difference and spatial
distribution overlap between the mixed modes. Moreover, we manifest that
the controllable JTE distortions at the sub-angstr\"o{}m level achieved
through partial isotope substitution can be unambiguously distinguished by
high-resolution TERS imaging.

\begin{figure}
\includegraphics[width=1.0\columnwidth]{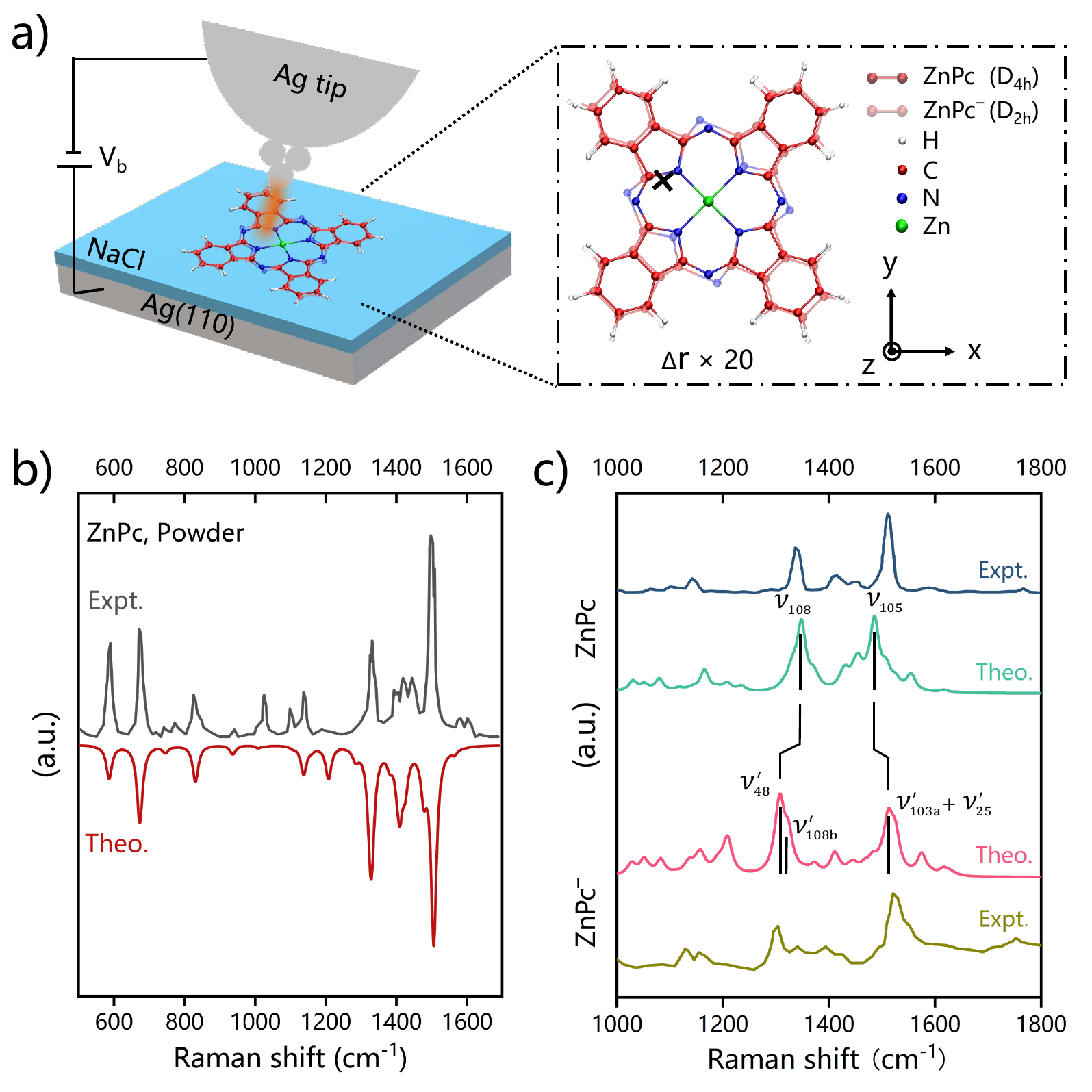}
\caption{(a) Schematic of experimental TERS measurement for
charge-induced JTE in a ZnPc molecule. The relative geometrical
differences between ZnPc and ZnPc$^{-}$ were amplified by a factor of
20 for a better illustration. The black cross mark indicated the
plasmon central position for the TERS spectra in (c). (b) Experimental
(gray line) and theoretical (red line) normal Raman spectra of ZnPc
powder. (c) Experimental and theoretical (under a 3~\AA{} plasmon) TERS
spectra of ZnPc and ZnPc$^{-}$  with plasmon center focused on the black
marked position in (a).  All the experimental spectra were extracted
from Ref.~\citenum{2023angew}. The label of vibrations
was in accordance with Herzberg's nomenclature\cite{Herzbergscheme} and
the superscript ``$\prime$'' represents the evolutionary counterparts of
the anionic form.}
\label{FIG.1}
\end{figure}

Figure~\ref{FIG.1}(a) illustrates the experimental setup for injecting
an electron into neutral ZnPc to induce JTE distortion. Previous
experiments observed sudden changes in TERS spectra when the sample bias
exceeded 0.3~V, which was assigned as a transition from the
neutral to negatively charged state\cite{2023angew}. It should be
stressed that owing to the insulating spacer layer introduced in
Fig.~\ref{FIG.1}(a), the negatively charged \ce{ZnPc-} has a
sufficiently long lifetime for spectral measurements and even
imaging\cite{2023angew}. Besides, the JTE-induced geometrical distortion
from the neutral $\text{D}_{\text{4h}}$ to the anionic 
$\text{D}_{\text{2h}}$ is quite subtle (the elongations along both the diagonal
axes are 0.026 and 0.010~\AA{}, respectively, as shown in
Fig.~\ref{FIG.1}(a)), which provides an ideal example for focusing
exclusively on vibrational variation itself.

To efficiently excite the required highly localized plasmonic field, an
incident light with a wavelength of 532~nm was employed in
experiments\cite{2023angew}. Under the same incident energy, our
first-principles calculations accurately reproduce the observed Raman
spectra of ZnPc powder under the experimental conditions
(Fig.~\ref{FIG.1}(b))\cite{2023angew}. Furthermore, by placing a 3~\AA{}
localized plasmon in the same position in experiments, i.e., above the lobe
of ZnPc (see the black cross in Fig.~\ref{FIG.1}(a)), the calculated 
TERS spectra for both neutral ZnPc and anionic ZnPc$^{-}$ obtained using the effective
field Hamiltonian\cite{duan2015jacs,duan2017jcp,2019xiejpcc,duan2019jacs,duan2024csr}
(also see Section~S1 in Supplemental Material) 
agree well with their experimental counterparts (Fig.~\ref{FIG.1}(c)).
Specifically, the calculated TERS spectrum of ZnPc$^{-}$ presents a 39.1~cm$^{-1}$
redshift and 25.8~cm$^{-1}$ blueshift compared with the two dominant bands
around 1348 and 1486~cm$^{-1}$ in ZnPc, respectively, which consistently
reproduces the correspondingly observed 45~cm$^{-1}$ redshift and
22~cm$^{-1}$ blueshift in experiments\cite{2023angew}. The
quantitative reproduction of the measured references in both uniform
and plasmonic fields validates the accurate descriptions of
electronic and vibrational structures as well as their responses at the
current theoretical level.

According to the calculations, the experimentally measured\cite{2023angew} two 
dominant bands for ZnPc and ZnPc$^{-}$ (Fig.~\ref{FIG.1}(c)) are assigned to 
different vibrational modes. Specifically, the degenerate $\nu_{\text{108}}$ 
and $\nu_{\text{105}}$ contribute to the dominant bands of ZnPc. In ZnPc$^{-}$, these
vibrations split as a consequence of JTE and are labeled with subscripts
``$\text{a}$'' and ``$\text{b}$'' hereafter. As a result, the dominant
high-frequency band originates from the occasionally degenerate 
$\nu_{\text{103a}}^{\prime}$ and $\nu_{\text{25}}^{\prime}$, where the
superscript ``$\prime$'' denotes the anionic form. Meanwhile, 
$\nu_{\text{48}}^{\prime}$ contributes to the  dominant band with low
frequency, with the shoulder assigned as $\nu_{\text{108b}}^{\prime}$.
Although this assignment is different from previous reports\cite{2023angew},
the observation of only a single splitting component, i.e.,
$\nu_{\text{108b}}^{\prime}$ in ZnPc$^{-}$, indicates the spatial
localization of degenerate vibrational modes during symmetry breaking
induced by JTE.

\begin{figure*}
\includegraphics[width=1.8\columnwidth]{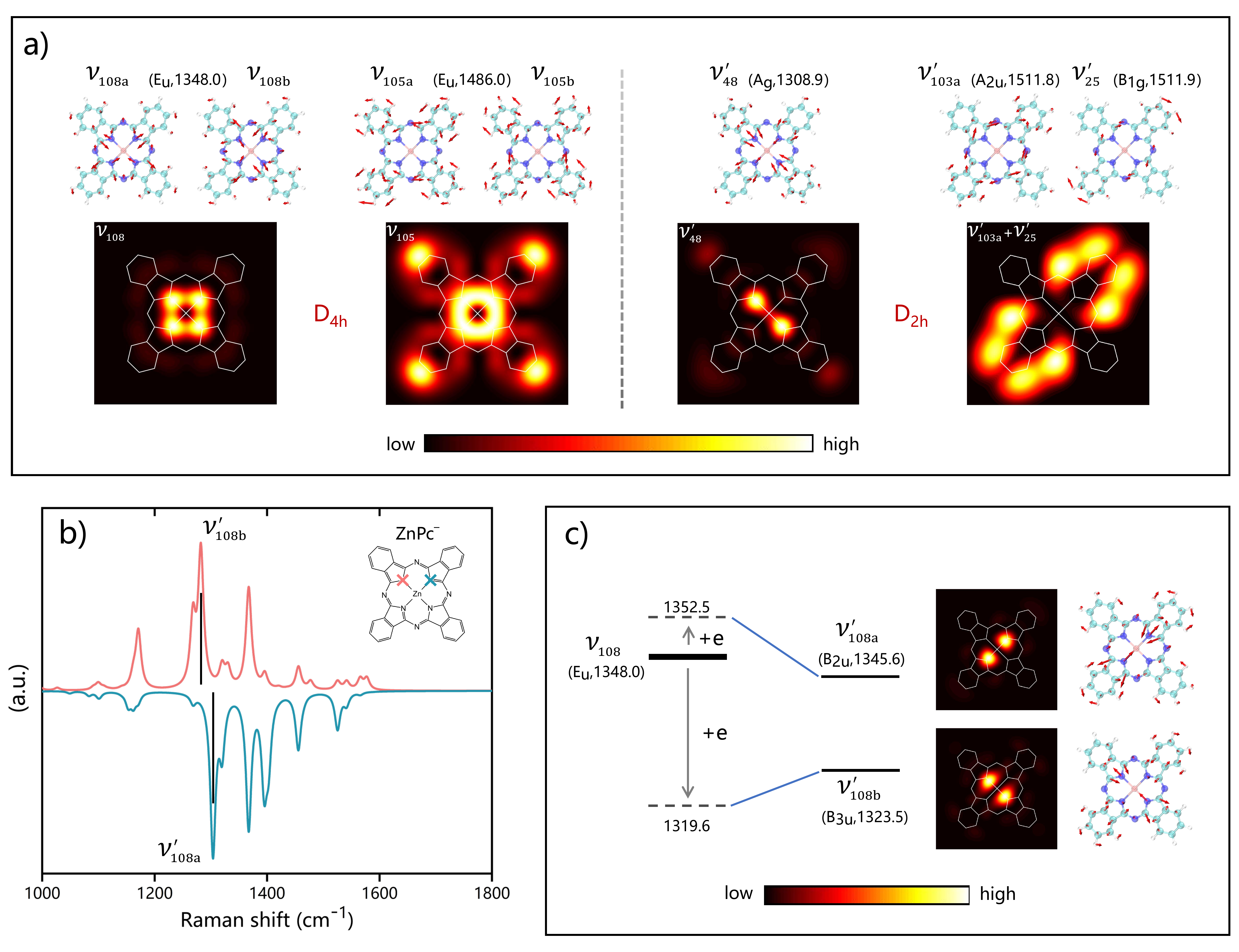}
\caption{(a) Calculated Raman images under a 3~\AA{} plasmon for the
dominant bands of ZnPc (left) and ZnPc$^{-}$ (right) as shown in
Fig.~\ref{FIG.1}(c).  The corresponding vibrational modes, frequencies,
and irreducible representations were depicted for comparison.  (b)
Calculated TERS spectra under a 3~\AA{} plasmon located at nitrogen
atoms of different axes in ZnPc$^{-}$ (marked by crosses).  (c)
Illustration of the splitting of the degenerate vibrational mode
$\nu_{\text{108}}$ due to charge-induced JTE. The electronic effect of
the excess electron on the vibrational energy of $\nu_{\text{108}}$ was
indicated by arrows.  The calculated Raman images and corresponding
vibrational modes for the splitting modes in ZnPc$^{-}$ were depicted.}
\label{FIG.2}
\end{figure*}

To directly visualize the localization, we turn to vibrationally resolved TERS
imaging\cite{2015duanAngew,zhang2019,lee2019nature,xz2023JPCL,yaolun2024jpcl}.
By moving tip positions and simultaneously collecting scattering
intensities at a given Raman shift, the images of dominant bands of
\ce{ZnPc} were calculated as shown in Fig.~\ref{FIG.2}(a).  Both images
exhibit clear four-fold symmetrical patterns, which correspond to the
$\text{E}_{\text{u}}$ irreducible representation in $\text{D}_{\text{4h}}$
point group (see vibrational displacements in Fig.~\ref{FIG.2}(a)).
On the other hand, the dominant modes of ZnPc$^{-}$ give 
two-fold symmetric patterns (Fig.~\ref{FIG.2}(a)), demonstrating the symmetry
reduction from $\text{D}_{\text{4h}}$ to $\text{D}_{\text{2h}}$ owing to
JTE. The individual contributions of component modes to images in  
Fig.~\ref{FIG.2}(a) can be found in Figs.~S1-S3 (Section~S2).

Detailed analysis reveals that at the central Zn position, the image of
$\nu_{\text{48}}^{\prime}$ is constructive, which is opposite to the
destructive feature in that of $\nu_{\text{108}}$. This result further
confirms that the dominant bands observed in experiments\cite{2023angew}
originate from different vibrational modes. To determine the split modes of
degenerate vibrations, for example, $\nu_{\text{108}}$ that involves vibrations
at four interior nitrogen atoms, we focused the plasmon field on the
nitrogen atoms along different axes in ZnPc$^{-}$ (see the color-marked crosses in the inset
of Fig.~\ref{FIG.2}(b)). The calculated spectra manifest that
one split mode becomes active, while another one is completely suppressed
(Fig.~\ref{FIG.2}(b)). As a result, the split modes ($\nu_{\text{108a}}^\prime$
and $\nu_{\text{108b}}^\prime$) can be identified. TERS images of these
vibrations exhibit split patterns derived from the original
$\nu_{\text{108}}$, with a destructive center, as anticipated (Fig.~\ref{FIG.2}(c)).

With the correct assignment, the electronic and displacement
redistribution effects can be unambiguously investigated for the splitting of
degenerate vibrations when the molecule experiences a JTE distortion.
Particularly, adding an extra electron to neutral $\text{D}_{\text{4h}}$
ZnPc would split the degenerate $\nu_{\text{108}}$ at 1348.0~cm$^{-1}$
into a blue-shifted mode at 1352.5~cm$^{-1}$ and a significantly
red-shifted mode at 1319.6~cm$^{-1}$. The $\text{E}_{\text{u}}$
$\nu_{\text{108}}$ mode then becomes further localized, which results in two
non-degenerate modes, i.e., a $\text{B}_{\text{2u}}$ mode
($\nu_{\text{108a}}^{\prime}$) at 1345.6~cm$^{-1}$ and a 
$\text{B}_{\text{3u}}$ mode ($\nu_{\text{108b}}^{\prime}$)
at 1323.5~cm$^{-1}$ (Fig.~\ref{FIG.2}(c)). The decomposition of
electronic and vibrational effects is a general approach. The detailed analysis
for other representative degenerate modes can be found in Fig.~S4. It should be
stressed that the minor geometrical distortion caused by JTE in ZnPc 
enables the decomposition of these effects.

\begin{figure*}
\includegraphics[width=1.8\columnwidth]{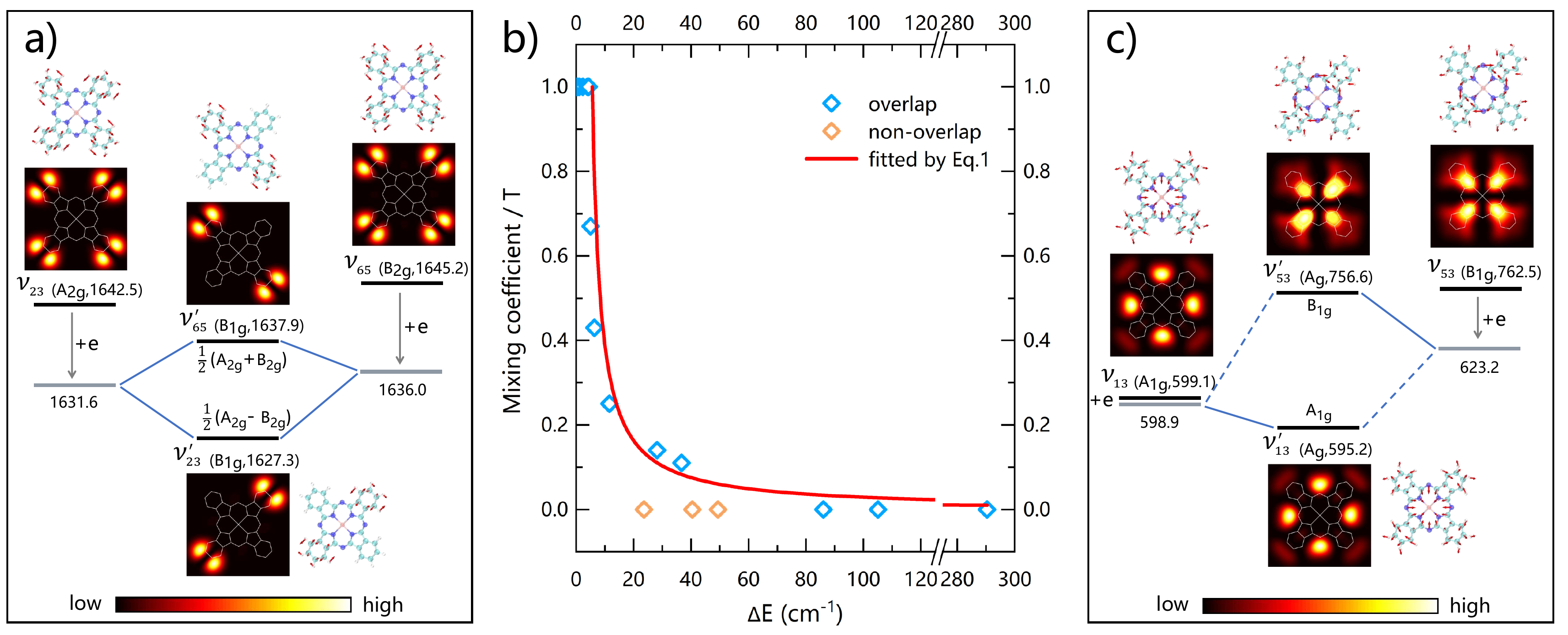}
\caption{Evolution of non-degenerate vibrational mixing from ZnPc to
ZnPc$^{-}$. (a) Mixing of well-overlapped
$\nu_{\text{23}}$ and $\nu_{\text{65}}$ modes.  (b) Relationship
between the mixing coefficient $T$ and the energy difference $\Delta E$ for
mixed modes, fitted by Eq.~\ref{equ.1} (the red line) and measured by
Raman images (the diamond scatters).  (c) Mixing of non-overlapped
$\nu_{\text{13}}$ and $\nu_{\text{53}}$ modes.}
\label{FIG.3}
\end{figure*}

The minor geometrical distortion also allows explicit visualization of
another vibrational evolution, i.e., the mixing of non-degenerate
vibrations in high symmetry after experiencing JTE (Table~S1), a phenomenon
completely overlooked in previous studies.
For instance, the $\text{A}_{\text{2g}}$-symmetric $\nu_{\text{23}}$ and
$\text{B}_{\text{2g}}$-symmetric $\nu_{\text{65}}$ in ZnPc would mix
with each other, generating two $\text{B}_{\text{1g}}$ modes 
($\nu_{\text{23}}^{\prime}$ and $\nu_{\text{65}}^{\prime}$) in ZnPc$^{-}$
(Fig.~\ref{FIG.3}(a)). A detailed decomposition reveals that the
electronic effect leads to redshifts of two modes, resulting in a
frequency difference of 4.4~cm$^{-1}$. It is noted that 
the displacements of $\nu_{\text{23}}$ and $\nu_{\text{65}}$ are all
located on the lobes with an efficient overlap. Calculated induced dipole
moments (Figs.~S5 and S6) determine that the mixing coefficient (see definition in Eq.~\ref{equ.1}) 
between $\nu_{\text{23}}$ and $\nu_{\text{65}}$ is close to 1.0, which results
in distinctly split patterns in $\nu_{\text{23}}^\prime$ and
$\nu_{\text{65}}^\prime$ as shown in Fig.~\ref{FIG.3}(a).

The mixing coefficient could be analogously derived from the
perturbational molecular orbital theory as (see a detailed derivation in
Section~S3) 
\begin{equation}
T=\frac{\lambda{S}_{12}}{\Delta{E}-\lambda{S}_{12}^2},
\label{equ.1}
\end{equation}
where $\lambda$ is a fitting parameter, $S_{12}$ is the overlap between two mixing
vibrations, and $\Delta{E}$ is the energy difference. For 
well-overlapped modes, i.e., $S_{12}$ close to 1, the mixing coefficient
$T$ is inversely proportional to the energy difference $\Delta{E}$,
which is exactly the case of mode mixing for $\nu_{\text{23}}$ and
$\nu_{\text{65}}$. Most of the mixing modes are in this category
(Figs.~S7-S9). By collecting all these modes, the fitting parameter
$\lambda$ is determined to be 2.8~cm$^{-1}$ (see the blue diamond
scatters and red line in Fig.~\ref{FIG.3}(b)).

The theoretical model also predicts another category with $S_{12}=0$,
where $T$ becomes zero regardless of $\Delta{E}$ and $\lambda$. In fact, there
are few modes belonging to this category. For instance, the electronic effect
leads to $\Delta{E}$ of 24.3~cm$^{-1}$ between the $\text{A}_{\text{1g}}$-symmetric $\nu_{\text{13}}$ and
$\text{B}_{\text{1g}}$-symmetric $\nu_{\text{53}}$ modes. If they were well-overlapped modes, the predicted $T$
would be 0.13. Nevertheless, with no mixing with each other, these two
modes give two $\text{A}_{\text{g}}$-symmetric modes in ZnPc$^-$,
exhibiting nearly identical four-fold image characteristics to their
respective counterparts in ZnPc (Fig.~\ref{FIG.3}(c)). 
This result should be attributed to the lack of
spatial overlap between $\nu_{\text{13}}$ and
$\nu_{\text{53}}$ modes, which is the sufficient prerequisite of $S_{12}=0$.
Two additional cases fall into this category (see orange diamond
scatters in Fig.~\ref{FIG.3}(b) and Fig.~S10).              

\begin{figure}[!tbh]
\includegraphics[width=1.0\columnwidth]{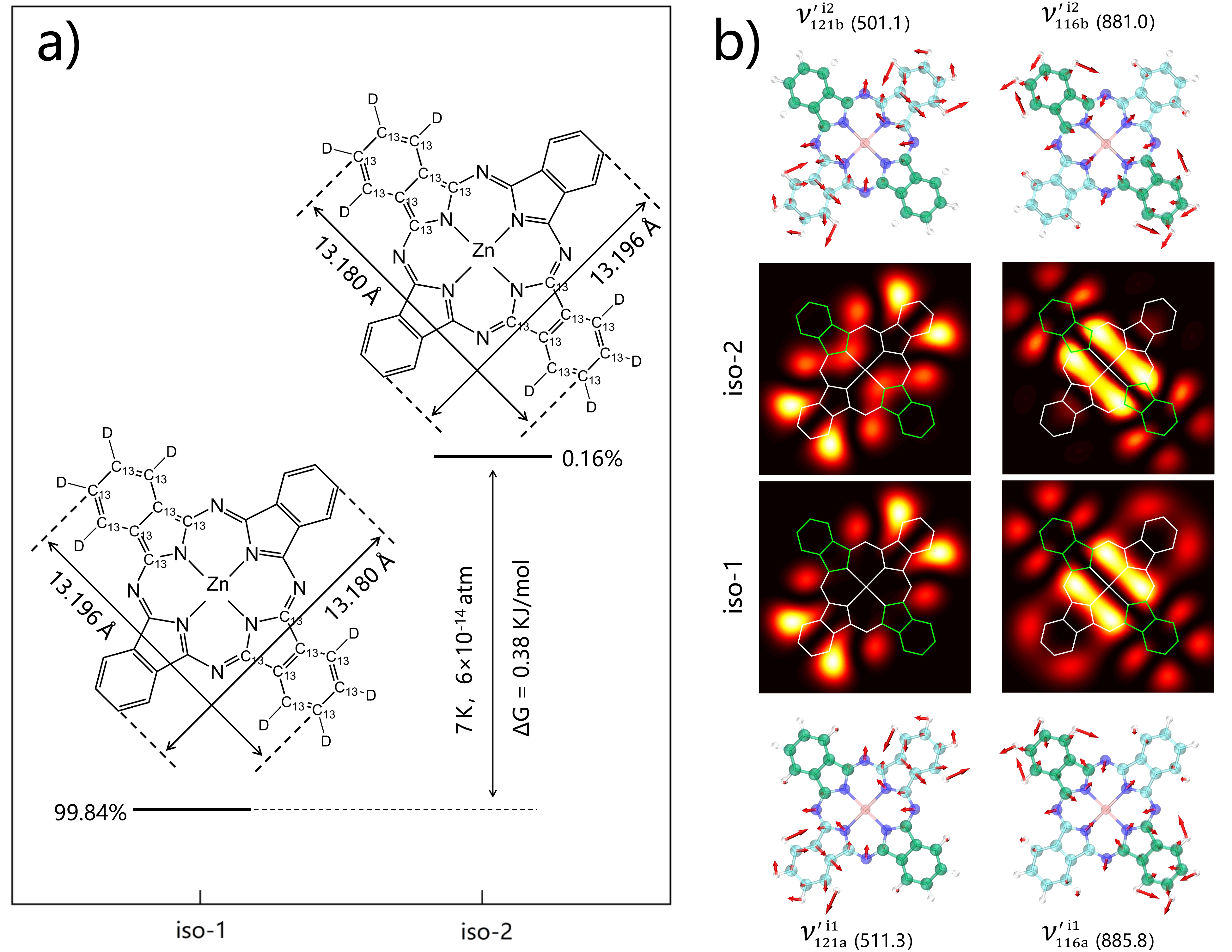}
\caption{Identification of JTE distortion in partially isotopic ZnPc$^-$.
(a) Two isomers (iso-1 and iso-2) of ZnPc$^{-}$ with partial isotope
substitutions of carbon-13 (${}^{\text{13}}\text{C}$) and deuterium (D)
in one diagonal axis. The labeled values are the calculated free energy
difference between the two isomers and their corresponding Boltzmann
populations under conditions of 7~K and $6\times10^{-14}$~atm.  (b)
Representative Raman images for analogical modes in iso-1 and iso-2.
The isotopic axis in vibrational modes and molecular skeletons
was highlighted in green.}
\label{FIG.4}
\end{figure}

The last capability of TERS imaging proposed in the present work is the
identification of the controllable distortion in JTE. To this end,
partial isotope substitution is employed to control the 
direction of JTE distortion.  Although the geometrical variation is quite small
(Fig.~\ref{FIG.1}(a)), isotope substitutions of ${}^{\text{13}}\text{C}$
and deuterium (D) in one diagonal axis can stabilize the ZnPc$^-$'s elongation
configuration along this axis, i.e., iso-1 (Fig.~\ref{FIG.4}(a)).
Calculated results reveal that at the experimental conditions of 7~K and
$6\times10^{-14}$~atm, iso-1 has a lower Gibbs free energy of 0.38~kJ/mol than
the configuration elongated along the other direction (iso-2). The
population of iso-1 is then determined to be 99.84\% in the experimental
conditions (Section~S4).  Because of the almost identical geometries
(Fig.~\ref{FIG.4}(a)), iso-1 and iso-2 are indistinguishable in
electronic structure-based SPM techniques. 

Remarkably, vibrationally resolved TERS images have the ability to overcome
this challenge (Fig.~\ref{FIG.4}(b)). Although the calculated TERS image of
$\nu_{\text{121a}}^{{\prime}\text{i1}}$ in iso-1 has the identical brightest lobe
patterns to that of the counterpart in iso-2, two moderate
patterns appear around the interior nitrogen atoms along the isotopic axis
for $\nu_{\text{121b}}^{{\prime}\text{i2}}$ in iso-2. Similarly, except for the
identical brightest central patterns, moderate patterns arise
around the non-isotopic lobes in $\nu_{\text{116a}}^{{\prime}\text{i1}}$
image for iso-1. These results should be attributed to the subtle
vibrational redistributions caused by the axial length difference at the
sub-angstr\"o{}m level. The ability of TERS imaging to distinguishing such sublet
vibrational variations offers a unique means for identifying
different configurations arising from isotopic JTE distortion.   

In summary, taking the experimentally feasible ZnPc as an example, we
demonstrate that vibrationally resolved Raman images have the capability
to monitor detailed vibrational evolutions when molecules experience
JTE distortions. The significance of identifying the most
common degenerate vibrational splitting in JTE via Raman images has been
highlighted. In addition, Raman images can further capture the mixing of
different irreducible representations in high-symmetry point groups due
to JTE. Moreover, we reveal that controllable JTE-induced distortion
configurations, achieved through partial isotopic substitution, can also be
distinguished via Raman imaging. These findings manifest that
vibrationally resolved Raman imaging provides a unique and practical
optical means for comprehensively monitoring vibrational evolutions in
JTE, deepening our understanding and enabling the regulation of JTE.

\begin{acknowledgments}
This work was supported by the National Key R\&D Program of China
(2024YFA1208104), the National Natural Science Foundation of China (Nos.
22393911, 22373060, 22473028, and 12474258), the Innovation Program for Quantum
Science and Technology (2021ZD0303301), Shandong Provincial Natural
Science Foundation (projects ZR2021QB164 and ZR2024QA020), and the
Taishan Scholar Project of Shandong Province (tsqn202211110).
\end{acknowledgments}

\bibliography{JTE.bib}
%\end{thebibliography}%

\end{document}